\documentclass[aip,jcp,numerical,reprint]{revtex4-1}

\usepackage{amsmath,amssymb}
\usepackage{graphicx}
\usepackage{dcolumn}
\usepackage{bm}
\usepackage{algorithm}
\usepackage{color}

\begin{document}

\preprint{APS/123-QED}

\title{A universal preconditioner for simulating condensed phase
  materials}

\author{David Packwood}
\affiliation{%
  Mathematics Institute, University of Warwick, Coventry CV4 7AL, United Kingdom
}%

\author{James Kermode}
\email{j.r.kermode@warwick.ac.uk}
\affiliation{%
  Warwick Centre for Predictive Modelling, School of Engineering, University of Warwick, Coventry CV4 7AL, United Kingdom
}%

\author{Letif Mones}
\affiliation{
Engineering Laboratory, University of Cambridge, Trumpington Street, Cambridge, CB2 1PZ, United Kingdom
}%
\affiliation{%
  Mathematics Institute, University of Warwick, Coventry CV4 7AL, United Kingdom
}%

\author{Noam Bernstein}
\affiliation{
Center for Materials Physics and Technology, Naval Research Laboratory, Washington DC 20375, USA
}

\author{John Woolley}
\affiliation{
 Department of Physics, University of Warwick, Coventry CV4 7AL, United Kingdom
}

\author{Nicholas Gould}
\affiliation{
Scientific Computing Department, STFC-Rutherford Appleton Laboratory Chilton, Oxfordshire OX11 0QX, United Kingdom
}

\author{Christoph Ortner}
\email{c.ortner@warwick.ac.uk}
\affiliation{%
  Mathematics Institute, University of Warwick, Coventry CV4 7AL, United Kingdom
}%

\author{G\'abor Cs\'anyi}
\email{gc121@cam.ac.uk}
\affiliation{
Engineering Laboratory, University of Cambridge, Trumpington Street, Cambridge, CB2 1PZ, United Kingdom
}%

\date{\today}

\begin{abstract}
  We introduce a universal sparse preconditioner that accelerates geometry
  optimisation and saddle point search tasks that are common in the atomic scale
  simulation of materials.  Our preconditioner is based on the neighbourhood
  structure and we demonstrate the gain in computational efficiency in a wide
  range of materials that include metals, insulators and molecular solids. The
  simple structure of the preconditioner means that the gains can be realised in
  practice not only when using expensive electronic structure models but also
  for fast empirical potentials.  Even for relatively small systems of a few
  hundred atoms, we observe speedups of a factor of two or more, and the gain
  grows with system size. An open source Python implementation within the
  Atomic Simulation Environment is available, offering interfaces to a wide
  range of atomistic codes.
\end{abstract}

\maketitle

\section{Introduction}

Geometry optimisation, i.e. finding a nearby local minimum of the potential
energy surface is the most common routine task of atomistic modelling, not only
used for finding the equilibrium geometries of molecules and crystals but also
as a fundamental building block of more complex algorithms for global
optimisation,\cite{Wales2003} structure prediction by random
search\cite{Pickard2011} and sampling.\cite{Voter1997} The closely related task
of finding saddle points is also used for finding transition states of
reactions, global optimisation, and accelerated sampling.

It is well recognised in the optimisation community how important
preconditioners are in creating efficient algorithms.  An example familiar in
the electronic structure community is using the kinetic energy operator as a
preconditioner when solving the electronic energy minimisation problem in plane
wave pseudopotential density functional theory (DFT) codes.\cite{Payne1992}
Preconditioning in linear algebra and numerical PDE problems is well
established, but ``universal'' preconditioners do not work particularly well,
and most practitioners advocate constructing preconditioners specifically
designed to suit each problem.\cite{Elman2014} There is a middle ground, which
is to reduce the domain enough to be able to give a good preconditioner, but
keep it general enough that many problems that need solving fall into it.

The hallmark of a good preconditioner is that it captures some aspects of the
local curvature of the potential energy landscape, e.g. some of the directions
in which the minimum is much shallower than in other directions. In this way,
using the preconditioner enhances the convergence by reducing the condition
number (see \eqref{eq:kappa}). For example, it was recognised by many that
geometry optimisation with a computationally expensive electronic structure
model can be preconditioned using cheap empirical interatomic model. This
approach is clearly not feasible for large scale problems in which the modeling
method itself is a relatively cheap interatomic model.

A universal goal in preconditioning of condensed phase atomistic systems is to take account of the long wavelength vibrational modes, whose energies tend towards zero as the system size increases, while the eigenvalues corresponding to
the high frequency optical modes stay constant. In order to capture this geometry, due to the intrinsic locality of the interaction Hamiltonian, it is enough to build a model that is aware of the neighbourhood structure of the constituent atoms or molecules.

In this work we use the simplest preconditioner that is capable of capturing
this structure, the adjacency matrix of the atoms, or a smoothed variant using a
distance cutoff. The only requirement of the cutoff is that it is chosen such
that all atoms are assigned some neighbours.  We choose example systems of
current interest which have a wide range of system sizes.

For a steepest descent (SD) or nonlinear conjugate gradient (CG) scheme with
preconditioner $P$ one expects that the number $n_P$ of iterations required to
reach a relative residual $\tau$ is\cite{Nocedal2006}
\begin{equation}
  n_P \sim |\log\tau| \times \begin{cases} \kappa_P & \text{(SD)} \\
    \sqrt{\kappa_P} & \text{(CG)},
  \end{cases}
\end{equation}
where $\kappa_P$ is the condition number  of the preconditioned Hessian at
equilibrium,
\begin{equation}
\kappa_P = \lambda_\mathrm{ max}/\lambda_\mathrm {min},
\label{eq:kappa}
\end{equation}
and
\begin{eqnarray}
\lambda_\mathrm{max} &=& \max_{u} \frac{u^T Hu}{u^TPu}\\
\lambda_\mathrm{min} &=& \min_u \frac{u^THu}{u^TPu}
\label{eq:lammin}
\end{eqnarray}
are the largest and smallest eigenvalues.

For a material system with a diameter of $R$ atomic spacings, without preconditioning (i.e. $P\equiv I$), one
expects $\kappa_I \sim R$ while our preconditioner achieves that $\kappa_P$ is
independent of $R$. Therefore the expected efficiency gain is
\begin{equation}
  \frac{n_P}{n_I} \approx \begin{cases} R^{-1} & \text{(SD)} \\
    R^{-1/2} & \text{(CG)}. \label{eq:efficiency}
  \end{cases}
\end{equation}
The theory of the most commonly used Broyden-Fletcher-Goldfarb-Shanno (BFGS) and similar quasi-Newton type schemes is less clear, but
numerical evidence suggests that a similar conclusion as in the CG case can be
drawn.

\section{Methods}

\subsection{Geometry Optimisation}
Throughout, we let $f(x)$ denote the energy for a configuration $x$. If $x_k$ is
an iterate of an optimization algorithm then we denote the gradient and Hessian at $x_k$, by $g_k = \nabla f(x_k)$ and $H_k = \nabla^2 f(x_k)$, respectively.

The most basic geometry optimisation schemes are steepest descent and
(undamped) Newton's method,
\begin{align}
  \label{eq:gd}
  x_{k+1} &= x_k - \alpha_k g_k, \\
  \label{eq:newton}
  x_{k+1} &= x_k - H_k^{-1} g_k.
\end{align}
While the former suffers from slow convergence to equilibrium due to
ill-conditioning of the energy-landscape, the latter is usually impractical
since (i) analytical Hessians are typically unavailable  for complex interatomic
potentials and electronic structure methods and (ii) are expensive to invert.

Line search is an essential part of all the above gradient descent algorithms, and  preconditioning  the line search (as opposed to preconditioning  the Newton
step) can be thought of as a middle ground, replacing $H_k$ with an approximate
Hessian $P_k$,
\begin{equation}
  \label{eq:pgd}
  x_{k+1} = x_k - \alpha_k P_k^{-1} g_k.
\end{equation}
The usual requirements on $P_k$ are that it is (1) cheap to build; (2) cheap to
invert; and (3) positive definite to ensure descent in energy.

The most common way to construct $P_k$ is via a quasi-Newton approach, typically
(L)BFGS. This works poorly for large systems since many iterations are required
to ``learn the Hessian'' to a useful degree of accuracy.  {Physical
  intuition and mathematical analysis can been used to develop an improved
  initial guess for the Hessian to speed up convergence.\cite{Pfrommer1997} }

An alternative approach (sometimes used in the electronic structure community
\cite{Soler}) is to take $P_k = \nabla^2 \tilde{f}(x_k)$ to be the Hessian of a
surrogate interatomic potential model $\tilde{f}$. This has considerable
potential for performance gains if a good surrogate model $\tilde{f}$ can be
found. Downsides of this approach are (i) the challenge of finding or
constructing such a surrogate model; (ii) indefiniteness of the surrogate
Hessian in the nonlinear regime (and potentially even in the asymptotic regime);
(iii) lack of transferability of the preconditioner: changing the system
requires the construction of a new surrogate model.

\subsection{Metric preconditioning}
\label{sec:metric-preconditioning}

Assume, for the moment, that we use  the same preconditioner throughout the optimization process,  $P_k \equiv P$ in (\ref{eq:pgd}).  An alternative
point of view, which is common in the numerical linear algebra and nonlinear
optimisation communities, is to think of $P$ as defining a metric on the space
of configurations. To see this note that calling $-g_k = - \nabla f(x_k)$ the
{\em direction of steepest descent} is with reference to the $\ell^2$-norm
$\| u \|_I := (\sum |u_i|^2)^{1/2}$ (where $u$ is a direction in configuration space). If we measure distances in configuration
space with respect to the $P$-norm, $\|u\|_P = (u^T P u)^{1/2}$, then the
direction of steepest descent becomes
\begin{equation}
  \arg\min_{\|u\|_P = 1} u^{T}\nabla f(x)\propto - P^{-1} \nabla f(x).
\end{equation}
That is, \eqref{eq:pgd} is the natural steepest descent scheme with respect to
the metric $P_k$.  The advantage of this point of view is that it frees us from
the constraint of aiming to approximate the Hessian. Instead we are now
searching for an alternative notion of distance in configuration space, which is
a more general concept and a fixed choice of metric may exist that is  suitable for a wide
range of atomistic systems.

Equivalently, we may think of (\ref{eq:pgd}) in terms of a change of
coordinates. Let $\tilde x_k := P^{1/2} x_k$, and $F(\tilde x) = f(P^{-1/2} \tilde x)$, then the
``standard'' gradient descent scheme $\tilde x_{k+1} = \tilde x_k - \alpha_k \nabla F(\tilde x_k)$ is
equivalent to (\ref{eq:pgd}).

Since $\nabla^2 F(\tilde x) = P^{-1/2} \nabla^2 f(P^{-1/2} \tilde x) P^{-1/2}$ it follows that
the rate of convergence $x_k \to x$ of (\ref{eq:pgd}) to a limit $x$ is
given by\cite{Bertsekas}
  \begin{displaymath}
    \|x_k - x\|_P \lesssim \big( {\textstyle \frac{\kappa_P - 1}{\kappa_P + 1} }
      \big)^k \| x_0 - x \|_P,
    \end{displaymath}
  where  $\kappa_P$ is the condition number of
  $P^{-1/2} H P^{-1/2}$.
  The latter can be computed from the generalised eigenvalue problem
\begin{equation}
  H v = \lambda P v.
\end{equation}

While approximating the Hessian would lead us to aim for $P$ such that
$\kappa_P \approx 1$, we shall be content with a good notion of distance which
will lead to a $P$ such that $\kappa_P$ is bounded by some moderate constant for a
wide range of systems of interest.

Our final remark in this abstract context is that while the discussion of
convergence rates applies strictly to the asymptotic regime of the iteration,
preconditioning also improves performance in the pre-asymptotic regime: a moderate upper bound
on $\kappa_P$ implies, loosely speaking, that (\ref{eq:pgd}) relaxes all
wavelength modes simultaneously rather than focusing on short wavelength modes
first.

\subsection{Preconditioned LBFGS}
\label{sec:lbfgs}
The usage of a preconditioner is not restricted to the steepest descent method,
but it can be readily applied to improved optimisation algorithms such as
nonlinear conjugate gradients. It is particularly effective when combined with
the LBFGS scheme \cite{Nocedal2006}, for which we briefly outline the
implementation.

Using $s_k = x_{k} - x_{k-1}$, $y_k = \nabla f(x_{k}) - \nabla f(x_{k-1})$,
$\rho_k = 1/y^T_k s_k$ then the action of the inverse Hessian can be efficiently
approximated,
\begin{align}
  \label{eq:LBFGSalg}
    \begin{split}
      &{\bf input}~q = \nabla f(x_k)  \\
      & {\bf output}~z \approx \nabla^2 f(x_k)^{-1} \nabla f(x_k) \\[2mm]
      &\mathrm{for } \: i = k,\ldots,k-m \\
      &\qquad \alpha_i = \rho_i s_i^T q \\
      &\qquad q = q - \alpha_i y_i \\
      &\fbox{$z = P_k^{-1} q$} \\
      &\mathrm{for } \: i = k-m,\ldots,k \\
      &\qquad \beta_i = \rho_i y_i^T z\\
      &\qquad z = z + (\alpha_i - \beta_i)s_i
    \end{split}
\end{align}
This formulation of LBFGS does not require the approximate Hessian itself to be
stored, only the positions and gradients at previous iterates. For the initial
iterate we simply obtain $z = P_0^{-1} \nabla f(x_0)$.  The boxed step is the
only modification needed to the standard algorithm to achieve preconditioning.
After obtaining the output $p_k =z$ from (\ref{eq:LBFGSalg}), the LBFGS step
takes the form
\begin{equation}
  \label{eq:LBFGS}
  x_{k+1} = x_k + \alpha_k p_k,
\end{equation}
for a suitable choice of step length $\alpha_k$.

\subsection{A simple and general metric for materials}
\label{sec:a simple metric}
Changes in energy of atomistic systems occur through changes in bonding, for which the simplest measure
is change in bond-length. Motivated by this observation we propose the following
preconditioner for materials systems: given parameters
$r_{\rm cut}, r_{\rm nn}, A, \mu$ (we will discuss below how to choose these
automatically) we define \(P\) via the quadratic form
\begin{align*}
  u^{T}P u   &= \mu \sum_{0 < |r_{ij}| < r_{\rm cut}} c_{ij} |u_i - u_j|^2, \\
  & c_{ij} = \exp\Big(- A \Big( {\textstyle \frac{r_{ij}}{r_{\rm nn}}} - 1 \Big) \Big)
\end{align*}
or, written in matrix form
\begin{align}
  \begin{split}
    P_{ij} & = \left\{ \begin{array}{rl} - \mu c_{ij}, & |r_{ij}| < r_\mathrm{cut}  \\
                         0, & |r_{ij}| \geq r_\mathrm{cut}  \end{array} \right. ,\\
    P_{ii} & = -\sum_{j\neq i} P_{ij}.
    \label{eq:C1}
  \end{split}
\end{align}
Default parameters are discussed in section \ref{sec:default}.

{\bf Remarks. }
(i) The exponential form of $c_{ij}$ is for convenience, and has no deeper
physical meaning; $A=0$ corresponds to using the adjacency matrix with a hard cutoff. (ii) We use this metric even for multi-component systems,
however, if the interaction strength and/or distances between different
components varies significantly, then it would be straightforward to generalise
it by distinguishing different types of bonds.  (iii) As shown in the Appendix,
for Bravais lattices, {\em phonon stability} is equivalent to the lower bound
$u^{T}H u \geq c u^{T} Pu $ for some constant $c > 0$.

Together with the generic and elementary upper bound $u^{T}Hu \leq C u^{T}Pu $
and equations \eqref{eq:kappa}-\eqref{eq:lammin} we obtain that for finite
periodic supercells in a Bravais lattice state, the condition number $\kappa_P$
for the preconditioned system is bounded above by $C/c$ independently of the
system size.  In the presence of defects (crystal surfaces, point defects,
dislocation lines) or even disorder partial results in this direction likely
still hold because \(P\)
contains the nearest neighbour bonds that dominate in \(H\).

\begin{figure}
  \includegraphics[width=0.54\textwidth]{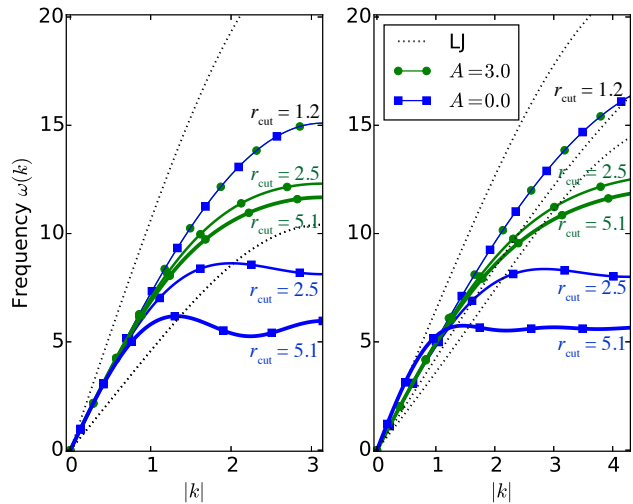}
  \caption{Spectra of the Lennard-Jones Hessian in the fcc ground state and of the
    preconditioner with $ A \in \{0, 3\}$ and
    $r_{\rm cut} \in \{1.2, 2.5, 5.1\} r_{\rm nn}$ (top to bottom, with increasing line thickness).
    The graphs for $A = 0, 3$ with $r_{\rm cut} = 1.2$ overlap.  Left panel: $(1,1,0)$
    direction, Right panel: $(2.7, 4.2, 3.0)$ direction.
    }
  \label{fig:spectra}
\end{figure}

\subsection{Default Parameters}
\label{sec:default}

The parameters $A$ and $r_{\rm cut}$ are user inputs, however $P$ is fairly
insensitive to their choice, provided their interdependency illustrated in
Figure \ref{fig:spectra} is taken into account. Hence, we suggest generic
default parameters below. The parameters $\mu$ and  $r_{\rm nn}$  are computed in a
preprocessing step from the initial configuration of the optimisation.

1. The nearest-neighbour distance $r_{\rm nn}$ is obtained as the maximum of
nearest neighbour bond-lengths: if $r_{\rm nn}^{(i)} = \min_{j \neq i} r_{ij}$
then $r_{\rm nn} = \max_i r_{\rm nn}^{(i)}$.

2. The exponent $A$ should be large enough to ensure that nearest neighbours
dominate, but not so large that small changes in the configuration lead to large
changes in $P$. All our tests are performed with $A = 0$ and $A = 3$; with
$A = 3$ giving slightly better performance.

3. The cut-off $r_{\rm cut}$ should be larger than $r_{\rm nn}$,
however, then exponential decay of the preconditioner entries ensures that
additional entries have a small influence. For \(A=0\) we choose
$r_{\rm cut} = 1.1 r_{\rm nn}$ and when $A=3$ we use $r_{\rm cut} = 2 r_{\rm nn}$. The latter choice is intuitively preferable since it accommodates
the possibility of significant bond stretching.

4. Finally, the energy-scale $\mu$ is chosen to ensure that the LBFGS algorithm
can choose the unit step-length as the default. We achieve this by equating
\begin{equation}
  v^{T}\big(\nabla E(x_0 + v) - \nabla E(x_0)\big)  = \mu v^T P_{\mu = 1} v, \label{eq:mu}
\end{equation}
where $P_{\mu=1}$ is the metric with $\mu = 1$ and $v$ is a test displacement of the
form
\begin{equation}
  v(x,y,z) = M \big(\sin(x / L_x), \sin(y / L_y), \sin(z / L_z)\big), \label{eq:perturb-mu}
\end{equation}
where $L_i$ are the lengths of the periodic lattice vectors and $M$ is a user-defined matrix with
default value $M = 10^{-2} r_{\rm nn} I$.

\subsection{Implementation details}
\label{sec:implementation details}
\textbf{Preconditioner application.} It is important that the cost of applying
the preconditioner does not dominate the cost of the calculation over the
evaluation of energy and gradient. For inexpensive models (Lennard-Jones, EAM,
Stillinger--Weber, etc) the choice of method to solve $z = P_k^{-1} q$ in
\eqref{eq:LBFGSalg} is crucial. Our implementation uses a smoothed aggregation
algebraic multigrid method \cite{PyAMG}. As a further optimisation we only
rebuild the preconditoner when the maximum atomic displacement since the last
update exceeds $r_{nn}/2$.

{\bf Line search.} Irrespective of the choice of the search direction used (e.g.,
SD (\ref{eq:pgd}), CG \cite{Nocedal2006} or LBFGS (\ref{eq:LBFGS})) a line
search algorithm must be implemented to choose the length of the step,
$\alpha_k$. The standard choice is a bracketing algorithm which enforces
sufficient decrease and approximate orthogonality between subsequent directions
(Wolfe conditions). We observed in our tests that a backtracking algorithm
imposing only sufficient decrease (Armijo condition), although less robust in
theory, was more efficient in practise. We give the details of our
implementation, and additional discussion, in Appendix~\ref{sec:linesearch}.

{\bf Robust energy differences. } The computation of the energy differences and
inner products in the Wolfe conditions (\ref{eq:Armijo}, \ref{eq:curv}) must be
performed with a high degree of accuracy, since the optimization algorithm
relies on robustly detecting the change in energy. A common difficulty in
implementing a line search strategy based on (\ref{eq:Armijo}, \ref{eq:curv}) is the
numerical round-off error that arises for large numbers of atoms (typically $10^5$ or
higher). Numerically robust inner products are equally important in the
inversion of the preconditioner and in the LBFGS algorithm. Numerically robust
evaluation of energy differences and inner products may, for example, be implemented using
compensated summation algorithms.\cite{HighamAccuracy} A simpler strategy which
proved sufficient in our case is to use 128~bit floating point numbers for these
steps.

{\bf Stabilisation. } If the system contains clamped atoms then the
preconditioner defined in \eqref{eq:C1} is strictly positive definite but in
order to improve its conditioning and ensure positive definiteness for cases where there
are no clamped atoms, we stabilize the preconditioner by adding a diagonal term,
\begin{align}
\begin{split}
    P_{ij} & = \left\{ \begin{array}{cc} - \mu c_{ij} & |r_{ij}| < r_\mathrm{cut}  \\  0 & |r_{ij}| \geq r_\mathrm{cut}  \end{array} \right. ,\\
    P_{ii} & = -\sum_{j\neq i}P_{ij} + \mu C_{\mathrm{stab}}.
\label{eq:C1stab}
\end{split}
\end{align}
In all our results we choose $C_\mathrm{stab}=0.1$.
Even when there are clamped atoms, we find that setting $C_\mathrm{stab} = 0.1$ improves
overall performance.

{\bf Variable cell optimisation. } We confirmed that our preconditoner also
gives good performance when degrees of freedom associated with the periodic unit cell
are included as well as the atomic positions.
Following the approach of Tadmor et al.~\cite{Tadmor1999}, we consider
a combined objective function $\Phi(x,D) = f(Dx)$ with $3N +9$ degrees of freedom:
$3N$ for the atomic positions $x$
and  9 components of the deformation tensor $D$, which is with respect to the original undeformed
unit cell.
The combined gradient is then given by
\begin{equation}
    \nabla_{x,D}\,\Phi(x, D) = \left( D \nabla_x f(x), \;
    \frac{V}{\mu_c} \sigma \left(D^{-1}\right)^T \right)
\end{equation}
where $V$ is the cell volume and  $\sigma$ the stress tensor, and we have
introduced an additional preconditioner
parameter $\mu_c$ to set the energy scale for the cell degrees of freedom.
$\mu_c$ can be pre-computed
at the same time as $\mu$ for no additional cost by including a trial
perturbation of the cell in \eqref{eq:perturb-mu},  with default
$v(x_c) = M/r_{nn} = 10^{-2} I$.

\section{Results}

\begin{figure}
  \centering
  \includegraphics[width=0.5\textwidth]{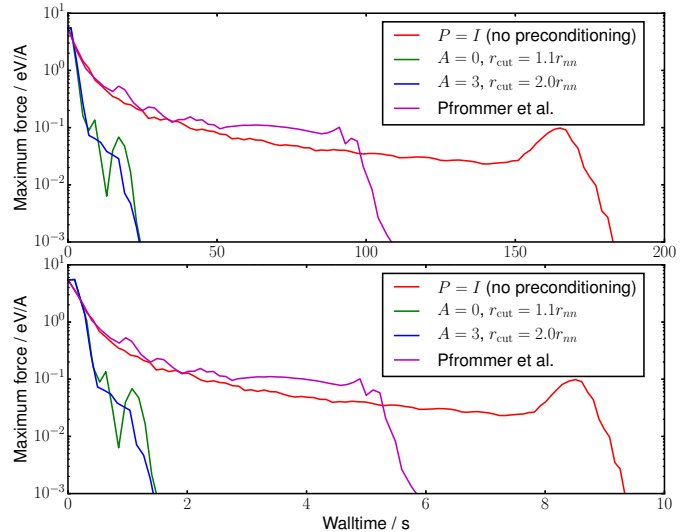}
  \caption{Convergence of the geometry optimisation of a 160-atom silicon slab using the Stillinger-Weber potential in fixed unit cell.
The parameters of the preconditioner
(or not using a preconditioner)\ are given in the legend. The lower panel shows the time required to
solve the problem in each case, indicating that the overhead of constructing and applying the preconditioner is minimal in comparison to the cost of computing forces with the interatomic potential.
\label{fig:si-slab}
  }
\end{figure}

We have selected a broad range of materials examples to test our preconditioner.
The first is a 160 Si atom $1\times1\times20$ supercell of the cubic diamond
structure cell in a slab geometry, with periodic boundary conditions along $x$
and $y$ and free boundaries in $z$, simulated with the Stillinger-Weber
interatomic potential.\cite{Stillinger1985} The two halves of the cell (along
$z$) are uniformly displaced toward each other by 0.5~{\AA}, creating a large
but very localized strain in the center of the slab. The problem is
ill-conditioned because the initial strain is localized, but reaching the
relaxed geometry requires all the slab atoms to move out towards the free
surfaces.  As shown in Fig.~\ref{fig:si-slab}, both the $A=0$ and $A=3$
preconditioners dramatically reduce the computational cost of the minimization,
by a factor of about 6 compared to the non-preconditioned
minimizer. {Results using the \citet{Pfrommer1997} block-diagonal
  approximation to the initial inverse Hessian are also shown for comparison.}
Note that even for this relatively fast interatomic potential the computational
cost of applying the preconditioner is nearly negligible, so the reduction in
computational time is nearly equal to the reduction in number of energy
evaluations.

\begin{figure}
  \centering
  \includegraphics[width=0.5\textwidth]{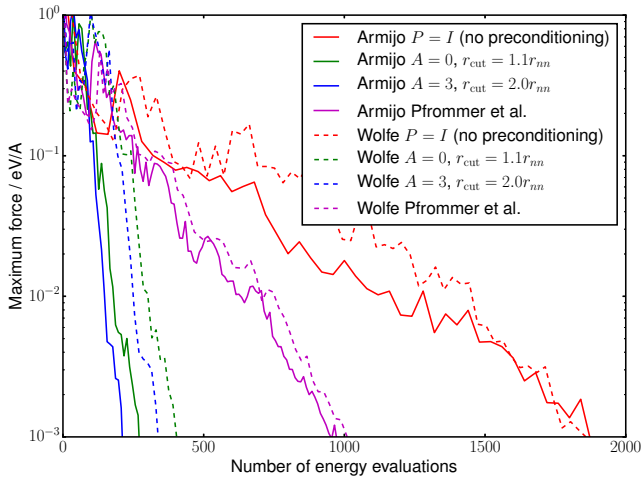}
  \caption{Convergence of the geometry optimisation of a silicon crack using the  Stillinger-Weber potential in a  fixed unit cell with 33,696 atoms. Solid and dashed lines correspond to using line searches enforcing Armijo and Wolfe conditions, respectively. The parameters of the preconditioner (or not using a preconditioner)\ are given in the legend.
  \label{fig:si-crack}
  }
\end{figure}

Next we consider a 33,696-atom Si model of the
$(111)[11\bar{2}]$ cleavage system (Fig.~\ref{fig:si-crack}) in a quasi-two-dimensional thin strip geometry with dimensions
$717\times242\times3.84$~\AA{}$^3$. The applied strain was chosen so that the
crack is lattice trapped~\cite{Thomson1971}, leading to a stable ground state
with the Stillinger-Weber~\cite{Stillinger1985} interatomic potential. Strong
coupling between length scales makes this a difficult system to optimize and
hence a good test of our preconditioner. A complex trade off between local
chemical cost and long-range elastic relaxation makes it favourable for a 5--7
crack tip reconstruction to form via a bond rotation.\cite{Kermode2008} Here, we
find that both the $A=0$ and $A=3$ preconditioners lead to a significant speed
up {over both unpreconditoned LBFGS and the approach of \citet{Pfrommer1997}.}
Fig.~\ref{fig:si-crack} also includes a comparison between the Armijo and
Wolfe line searches. As noted above, enforcing only the Armijo condition leads
to a further increase in performance.

\begin{figure}
  \centering
  \includegraphics[width=0.5\textwidth]{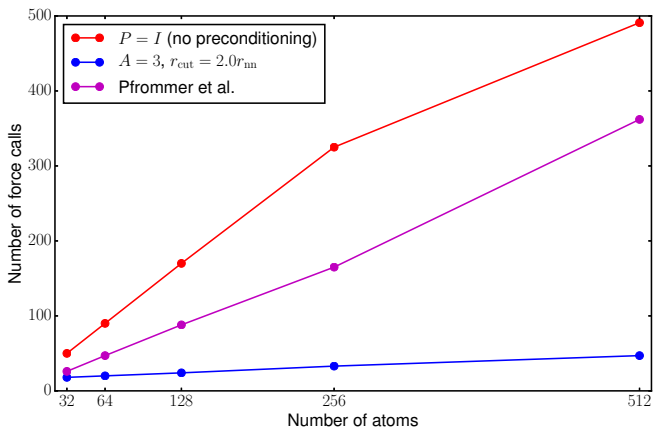}
  \caption{{Scaling with system size for geometry optimisations in $N\times1\times1$ Si supercells containing from 32 to 512 atoms with the Stillinger-Weber potential, using unpreconditioned LBFGS (red), the inverse Hessian approximation of Pfrommer et al. (magenta), and our new preconditioner (blue).
  \label{fig:precon-scaling}
  }}
\end{figure}

{To investigate whether the theoretical independence of the cost of
  preconditioned minimisations from system size (Eq.~\ref{eq:efficiency}) is
  achieved in real systems, we carried out tests in a series of
  $N\times1\times1$ Si supercells, again using the Stillinger-Weber
  potential. The atomic positions were perturbed by random displacements of
  magnitude 0.1~\AA{} and also subjected to a compressive strain of 0.5\% to
  introduce a long-wavelength deformation. The results shown in
  Fig.~\ref{fig:precon-scaling} indicate that our preconditioner achieves
  convergence after an approximately constant number of force evaluations as the
  system is made larger, in contrast to not using a preconditioner or to the
  approach of \citet{Pfrommer1997} which does not use connectivity
  information. Our new method is therefore expected to be particularily useful
  for very large systems.}

\begin{figure}
  \centering
  \includegraphics[width=0.5\textwidth]{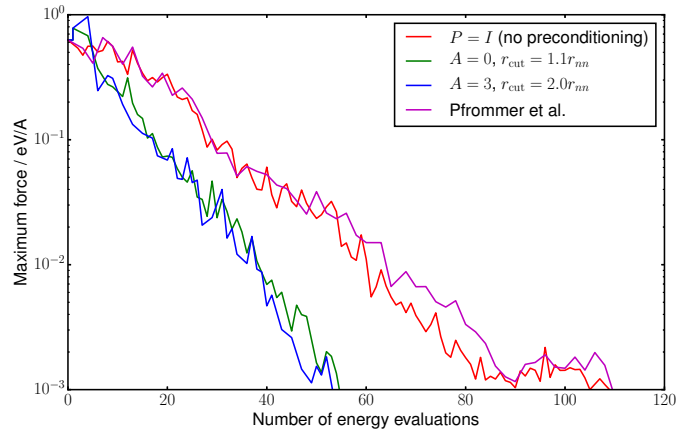}
  \caption{Convergence of the geometry optimisation of  a 220-atom  LaAlO$_3$ slab using  DFT in a fixed unit cell. The parameters of the preconditioner
(or not using a preconditioner)\ are given  in the legend.
  \label{fig:LaAlO3}
  }
\end{figure}

Since large systems inherently have a wide range of displacement
wavelengths and corresponding stiffnesses, it is not obvious {\it
a priori} how much preconditioning will help for a smaller system,
for example one that can feasibly be simulated using density
functional theory.  We therefore simulated a perovskite
structure oxide, LaAlO$_3$, in a 220-atom slab geometry with periodic
boundary conditions in-plane and free surfaces separated by a vacuum
region in the normal direction.  Energy and force evaluations used
DFT with the PBE exchange correlation functional, projector-augmented
waves (PAW) with a 282.8~eV cutoff plane-wave basis, and a
$2\times2\times1$ Monkhorst-Pack k-point sampling, evaluated using
the QUIP interface to the VASP software.\cite{Csanyi2007,Kresse1994,Kresse1996}
In this system, as shown in Fig.~\ref{fig:LaAlO3}, we find that the preconditioning still
significantly reduces the computational cost, but the improvement
is not as dramatic as for the larger systems discussed above.  With our
convergence criterion the reduction is about a factor of two,
although the non-preconditioned minimization stagnates just before
reaching convergence, and with a slightly looser criterion the
reduction would only be a factor of 1.6.  Note that the computational
cost of the DFT energy and force evaluations is so large that the
application of the preconditioner is completely negligible in
comparison. {In this case the approach of \citet{Pfrommer1997}
does not make a significant improvemenet over not using a preconditioner.}

\begin{figure}
  \centering
  \includegraphics[width=0.5\textwidth]{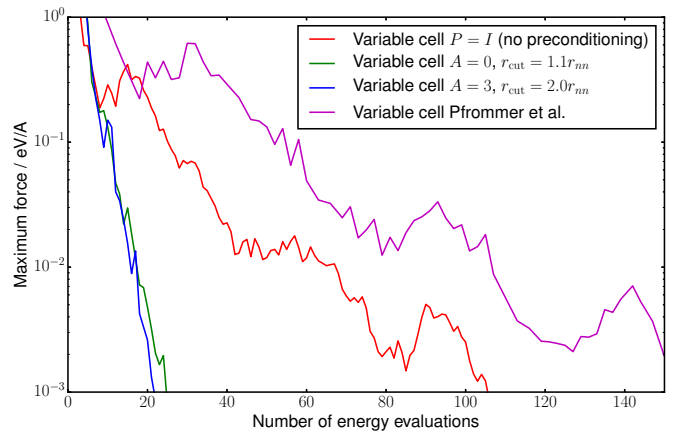}
  \caption{Convergence of the geometry optimisation of a 106-atom  $\gamma$-Al$_2$O$_3$ system in a variable cell. The parameters of the preconditioner
(or not using a preconditioner)\ are given in the legend.
  \label{fig:gamma-al2o3}
  }
\end{figure}

For a test of the relaxation of both atomic positions and unit cell size and
shape we used a $1 \times 1 \times 2$ supercell of a $\gamma$-Al$_2$O$_3$
structure, with methods similar to those described above for LaAlO$_3$, except
for a 530~eV plane wave cutoff and a $\Gamma$-centered k-point mesh.  For this
system, plotted in Fig.~\ref{fig:gamma-al2o3}, the reduction in computations for
both preconditioners is about a factor of 5, a very significant improvement.
While the non-preconditioned minimizer fails to make progress at several points
during the relaxation, both {our new} preconditioners allow the LBFGS
minimizer to rapidly and steadily reduce the gradient until convergence.
{Here, the approach of \citet{Pfrommer1997} actually results in slightly
  worse performance than unpreconditioned LBFGS.  This could perhaps be improved
  by careful tuning of the bulk modulus and optical phonon frequency parameters
  used to construct the approximation to the inverse Hessian; however, we note
  that our new preconditioner does not require any user input as all parameters
  are computed automatically.}  The addition of the cell degrees of freedom,
which are preconditioned in magnitude but not coupled to the positional degrees
of freedom, do not reduce the effectiveness of our preconditioners.

\begin{figure}
  \centering
  \includegraphics[width=0.5\textwidth]{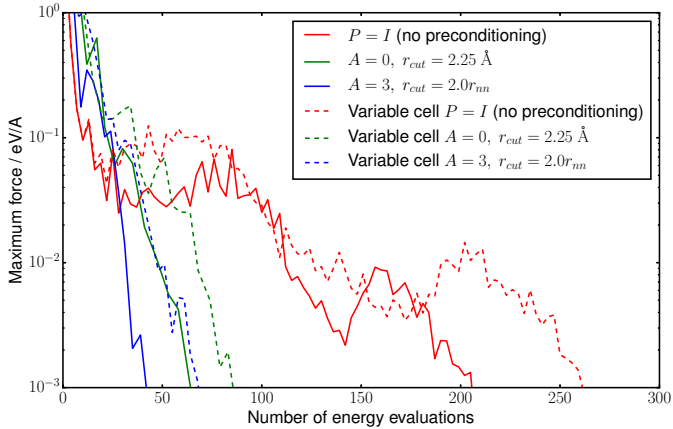}
  \caption{Convergence of the geometry optimisation of a 432-atom ice VIII system with fixed (solid lines) and variable (dashed lines) unit cells. The parameters of the preconditioner
(or not using a preconditioner)\ are given in the legend.
  \label{fig:ice}
  }
\end{figure}

Finally we tested the new preconditioner for a molecular system, ice VIII.
The system contained 432 atoms with an initial cell dimension of $13.65 \times 13.65 \times 13.16$ \r{A}$^{3}$.
A DFT potential with BLYP exchange-correlation functional was used with DZVP basis set and GTH
pseudopotentials. Calculations were performed by the CP2K program package using the QUIP interface.\cite{Csanyi2007,VandeVondele2005,Hutter2014}
Fig.~\ref{fig:ice} shows the number of energy evaluations of the different
optimisations for fixed and variable cells using a maximum force threshold of $10^{-3}$ eV \r{A}$^{-1}$.
Similarly to previous systems, the Armijo condition performed better than Wolfe
so we present here only the results with the former line search.
In the  $A=0$ case we slightly increased the default cutoff parameter ($r_{\rm{cut}}=2.25$ \r{A})
to include   hydrogen bonded neighbours too.
For both the fixed and variable cells the computational costs compared to the unpreconditioned
optimisation were reduced by 3 and 4 times using $A=0$ and $A=3$, respectively.

\section{Saddle search}
To demonstrate the transferability of our preconditioner not only across problem
classes but also across algorithms, we apply it to the dimer saddle search
algorithm.\cite{HenkJons:jcp1999, GPO-dimer} A modified variant of the
algorithm proposed in Ref.~\citealp{GPO-dimer} reads
\begin{align*}
  x_{k+1} &= x_k - \alpha \big( P_k^{-1} - 2v_k v_k^T \big)
            {\textstyle
            \frac{\nabla f(x_k+h v_k) + \nabla f(x_k-h v_k)}{2}}, \\
  v_{k+1}' &= v_k - \beta \big( I - P_kv_k v_k^T \big)  {\textstyle \frac{\nabla f(x_k + h v_k) - \nabla f(x_k - h v_k)}{2 h}}, \\
  v_{k+1} &= v_{k+1}' / \| v_{k+1}' \|_{P_{k+1}}.
\end{align*}
The translation step is obtained as coordinate transformation of the standard
dimer step with the variables
$\tilde{x}_k = P_k^{1/2} x_k, \tilde{v}_k = P_k^{1/2} v_k$ (cf. \S
\ref{sec:metric-preconditioning}).  The orientation step is an $\ell^2$-steepest
descent step (without preconditioning) for the Rayleigh-quotient
$v^T \nabla^2 f(x_k) v / v^T P_k v$, with a finite-difference approximation of
$\nabla^2 f(x_k) v$. Interestingly, naive preconditioning of the orientation
steps led to poorer performance in our tests.

\begin{figure}
  \centering
  \includegraphics[width=0.5\textwidth]{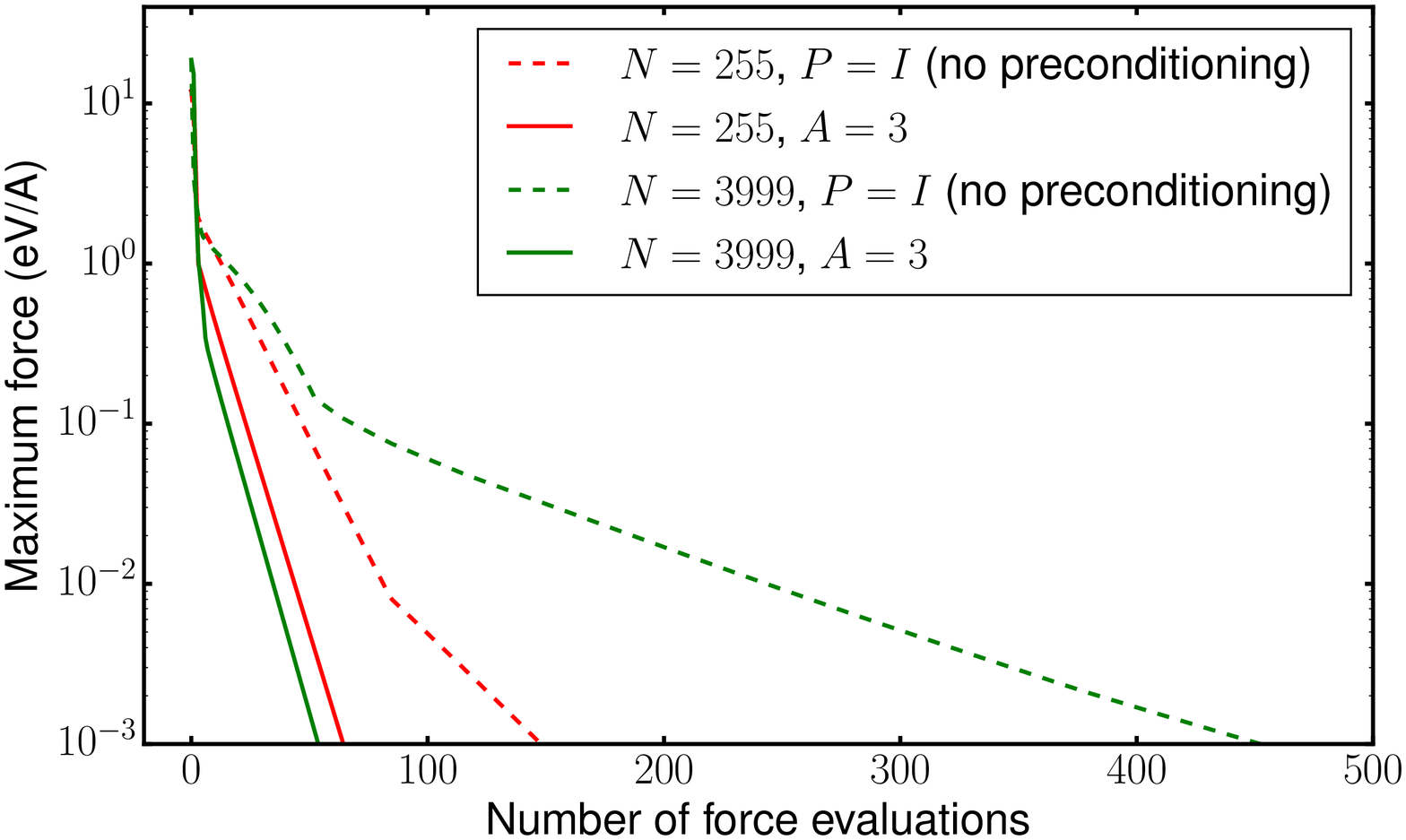}
  \caption{\label{fig:dimer} Performance of the preconditioned dimer method for
    a vacancy in a Lennard-Jones fcc crystal (solid lines), and without a
    preconditioner (dashed lines), for two system sizes.}
\end{figure}

We test this preconditioned dimer algorithm by computing the saddle
configuration of a vacancy in a Lennard-Jones fcc crystal, with a cubic
computational cell. Given two states $x^{(0)}, x^{(1)}$ which have two
neighbouring lattice sites removed, we choose the starting configuration
$x_0 = \frac13 x^{(0)} + \frac23 x^{(1)}$ and $v_0 \propto x^{(1)} - x^{(0)}$.
The step-sizes are chosen by hand-optimising for a small setup with $3^3$ unit
cells: $\alpha = 0.01, \beta = 0.005$ for the unpreconditioned variant
($P_k = I$) and $\alpha = 0.5, \beta = 0.01$ for our preconditioner with
parameters $A = 3.0, r_{\rm rcut} = 2 r_{\rm nn}$. For both variants we chose
$h = 10^{-2}$. The results are displayed in Figure~\ref{fig:dimer},
demonstrating analogous improvements to the energy minimisation examples.

\section{Conclusions}

 In summary, we have presented a simple preconditioner for geometry optimisation and saddle search that is universally applicable in a wide range of atomistic and molecular condensed phase systems, offering at least a factor of two in performance gain in our examples of small systems, and up to factor of ten in systems of tens of thousands of atoms. The extra cost of using the preconditioner is small enough that it is worth using even with inexpensive interatomic potentials, while the performance gain is expected to scale as the square root of the system size. A Python implementation within the Atomic Simulation Environment\cite{Bahn2002} is available at \url{https://gitlab.com/jameskermode/ase}, offering interfaces to a wide range of atomistic codes such as VASP\cite{Kresse1994}, CASTEP\cite{Clark2005}, CP2K\cite{VandeVondele2005}, LAMMPS\cite{Plimpton1995}, and many others.

\begin{acknowledgments}

This work was supported by the Engineering and Physical Sciences
Research Council (EPSRC) under grant numbers EP/J022055/1, EP/L014742/1,
EP/L027682/1, EP/J010847/1 and EP/J021377/1.
{An award of computer time was provided by the Innovative and Novel Computational Impact on Theory and Experiment (INCITE) program. This research used resources of the Argonne Leadership Computing Facility, which is a DOE Office of Science User Facility supported under Contract DE-AC02-06CH11357.}
Additional computing facilities were provided by the Centre for Scientific
Computing of the University of Warwick with support from the Science Research Investment Fund.
The work of N.~B. was supported by
the Office of Naval Research through the Naval Research Laboratory's basic
research program.
The work of C.~O. and L.~M. was also supported by ERC Starting Grant 335120.
We thank C.~S.  Hellberg and M.~D. Johannes for the LaAlO$_3$ and
$\gamma$-Al$_2$O$_3$ atomic configurations.

\end{acknowledgments}

\appendix

\section{Linesearch}
\label{sec:linesearch}
We present the details of our line search algorithm. The standard requirement
for the LBFGS and CG methods is that the step-size, $\alpha$, satisfies the
Wolfe conditions
\begin{align}
    \label{eq:Armijo}
      f(x_k+\alpha p_k) &\leq f(x_k) + c_1 \alpha \nabla f_k^T p_k, \\
    |\nabla f(x_k + \alpha p_k)^T p_k| &\leq c_2 |\nabla f_k^Tp_k|,
    \label{eq:curv}
\end{align}
where $0 < c_2 < c_1 < 1$. 
Line search methods that guarantee \eqref{eq:Armijo} and \eqref{eq:curv} employ a
bracketing strategy, which often requires several additional energy and force
evaluations at each iteration.

For the steepest descent method it is theoretically sufficient to impose only
the Armijo condition \eqref{eq:Armijo}. We have observed that this was also
sufficient in all our tests to ensure convergence of the LBFGS method and leads
to a consistent performance improvement. Our implementation minimises the
quadratic interpolating $f_k, \nabla f_k^T p_k$ and $f(x_k + \alpha p_k)$,
iterating until \eqref{eq:Armijo} is satisfied.  For $c_1 < 1/2$ this yields a
backtracking guarantee and hence ensures that the line search terminates
after finitely many steps. Our default parameter is $c_1 = 0.1$. The initial
estimate on the step-length is $\tilde\alpha = 1.0$.

\begin{align}
  \label{eq:linesearchALG}
  \begin{split}
    &{\bf input}~x, \tilde{\alpha} > 0, c_1 \in (0, 1/2), p
    \text{ s.t. } \nabla f(x) \cdot p < 0. \\
    &{\bf output}~\tilde{\alpha}. \\
    &\mathrm{while } \: f(x + \tilde\alpha p) > f(x)
    + c_1 \tilde\alpha \nabla f(x) \cdot p \\
    &\qquad \tilde\alpha' \leftarrow
    \frac{- \frac12 \tilde\alpha \nabla f(x) \cdot p}{
      \frac{f(x+\tilde\alpha p) - f(x)}{\tilde\alpha}
      - \nabla f(x) \cdot p} \\
    &\qquad \tilde{\alpha} \leftarrow \max(\tilde\alpha', \tilde\alpha/10)
  \end{split}
\end{align}

Unlike for a bracketing line-search, the only additional evaluations required
during line search are energy evaluations at the end-point of the search
interval, which reduces computational cost in the pre-asymptotic regime of the
optimisation.

In the asymptotic regime, the step-length $\alpha_k = 1$ is always accepted, and
will satisfy both Wolfe conditions \eqref{eq:Armijo} and \eqref{eq:curv}
provided that $c_1 < 1/2$. Since,
through the use of our proposed preconditioner, we substantially reduce the
number of iterations, it is unlikely that the {\em potential} instabilities
associated with Armijo line search for the LBFGS direction will be observed.
{Moreover, in our implementation, if the
Armijo linesearch fails we simply reset the LBFGS Hessian history and repeat
the linesearch, which removes any concern about robustness}.

\section{Phonon Stability}

Consider a $d$-dimensional Bravais lattice $\Lambda = A \mathbb{Z}^d$, where
$\mathbb{Z}$ is the set of integers and the columns of $A$ are the lattice
directions, which is the ground state for some material system under a potential
energy $f$. Let $H = \nabla^2 f$ denote the Hessian of the potential energy in
the ground state. For displacements $u_r$ of each atom $r \in \Lambda$ we can
write
\begin{equation}
  [Hu]_r = \sum_{s \in \Lambda} H_{rs} u_s,
\end{equation}
where $H_{rs} \in \mathbb{R}^{d \times d}$ are the blocks of $H$. We now prove
the claim that phonon stability is equivalent to the bound
$u^T H u \geq u^T P u$, where \(P\) is the preconditioner defined in~\eqref{eq:C1}, for all displacements of the lattice.

The discrete translation invariance of the lattice, $\Lambda + r = \Lambda$ for
all $r \in \Lambda$, implies that $H_{rs} = H_{0,s-r} =: h_{s-r}$ where
$h \in \mathbb{R}^{d \times d}$. For any virtual displacement $u = (u_r)$ with
compact support we have
\begin{equation}
  u^T H u = \int_{\rm BZ} \hat{u}^* \hat{h} \hat{u} \,dk,
\end{equation}
where \(\hat u \)
and \(\hat h\)
denote the Fourier transforms of \(u\)
and \(h\),
respectively and the integration is over the first Brillouin zone. Phonon
stability means that the natural frequencies are positive and linear near the
origin. In terms of $\hat{h}$, this translates to $\hat{h}(k) \geq c_H |k|^2 I$
for some constant $c_H > 0$. The upper bound $\hat{h}(k) \leq C_H |k|^2 I$
follows from the boundedness of the phonon band width. (A sufficient condition
is that $\sum_{r \in \Lambda} |h_r| |r|^2 < \infty$.)

Let $P_{rs} = p_{s-r} \in \mathbb{R}^{d \times d}$ denote the corresponding
blocks of the preconditioner operator. The upper bound
$\hat{p} \leq C_P |k|^2 I$ follows simply from the fact that the preconditioner
has a finite interaction range. This upper bound and phonon stability of $H$
imply
\begin{align*}
  u^T H u
  &\geq c_H \int_{\rm BZ} |k|^2 |\hat{u}|^2 dk \\
  &\geq \frac{c_H}{C_P} \int_{\rm BZ} \hat{u}^* \hat{p} \hat{u} \\
  &= \frac{c_H}{C_P} u^T P u.
\end{align*}

Conversely, if $u^T H u \geq c u^T P u$, then phonon stability of $P$ implies
phonon stability of $H$. But the former is an immediate consequence of the fact
that the coefficients in the definition of $P$ are positive.\cite{BornHuang}

\bibliography{precond.bib}

\end{document}